# Measurement of fluence, LET, and dose in a carbon ion spread-out Bragg-peak using fluorescent nuclear track detectors and an automated reader

Steffen Greilich[1,2,*], Leonie Ulrich[1,2], Jasper J.M. Kouwenberg[3], Shirin Rahmanian[1,2]

[1]*Division of Medical Physics in Radiation Oncology, German Cancer Research Center (DKFZ), D-69120 Heidelberg, Germany*

[2]*Heidelberg Institute of Radiation Oncology (HIRO), National Center for Radiation Research in Oncology (NCRO), D-69120 Heidelberg, Germany*

[3]*Radiation, Science & Technology, Delft University of Technology, Mekelweg 15, 2629 JB Delft, Netherlands*

**Abstract**

For the assessment of radiation effects of clinical ion-beams, dosimetry has to be complemented by information on particle-energy distribution or related quantities. Fluorescence nuclear track detectors made from C,Mg-doped alumina single crystals allow for the quantification of ion track density and energy loss on a single-track basis. In this study, their feasibility and accuracy to quantify fluence, linear-energy-transfer (LET) distributions, and eventually dose for a spread-out carbon ion Bragg peak was investigated. We found that the primary ions track densities agreed well with the reference data, but the determination of the individual detector sensitivity represented a major source of uncertainty in LET (and dose) assessment. While low-LET fragments in the beam are not contributing to this dose significantly, their number of was largely underestimated by approximately a factor three. The effect was most pronounced for protons where the measured fluence deviates at least an order of magnitude. We conclude that this is mainly caused by the wide angular distribution of protons in a carbon beam. The use of a dedicated FNTD reader device and semi-automated workflow improved outcome due to the considerably larger amount of data available as compared to a state-of-the-art multi-purpose confocal laser scanning microscope.

---

[*] Corresponding author: s.greilich@dkfz.de

**Introduction**

Ion Beam Cancer Therapy (IBCT), i.e. radiotherapy with heavy charged particles, is expected to show an improved clinical outcome for specific tumor types due to better dose conformity and less integral dose compared to conventional treatments using megavoltage X-ray or electron beams [1]. With increasing interest in IBCT, both the number of centers and treatments has surged during the last decade [2].

Important properties of ion beams, such as several quantities relevant for dosimetry, but especially their enhanced relative biological efficiency (RBE), depend on the 'radiation quality', i.e. the ionization density generated at a region of interest in the radiation field. The ionization density is mainly determined by the linear energy transfer (LET) $L = dE/dz$ of the individual particles, i.e. the energy spent in electronic collisions, $dE$, per unit path length $dz$ [3]. $L$ increases while the ions loose kinetic energy and slow down, but in addition clinical ion beams become inevitably multi-energetic, multi-particle beams by energy loss straggling and inelastic nuclear scattering. Thus, the particle spectrum, i.e. the fluence $d\Phi_Z(T)/dT$ differential in kinetic energy $T$ of ions with charge $Z$, and consequently the LET-spectrum $\Phi_L = d\Phi(L(T,Z))/dL$, change considerably when passing through tissue. Often, radiation quality is collapsed into a single number specifier like the fluence- or dose-averaged LET [4] or the residual range. $\Phi_L$ is closely related to microsdosimetric distributions of specific dose or lineal energy.

The gold standard for dosimetry in clinical ion beams are ionometric measurements using the charge produces in air-filled chambers. Current protocols such as the IAEA TRS-398 [5] use a quality correction factor $k_{Q,Q_0}$ to transfer measurements from one radiation quality, $Q_0$, to another, $Q$. $k_{Q,Q_0}$ represents usually a small correction: for a standard Farmer chamber (PTW30001) it does not change more than 5 ‰ in proton beams with residual ranges between 0.5 and 30 g/cm². This makes air-filled ionization chambers a powerful tool for high-accuracy dosimetry in ion beams but indicates at the same time that they can provide only very limited information on radiation quality.

Most solid-state dosimeters such as films, TLDs, OSLDs, alanine, gels, etc., suffer significant sensitivity dependence on radiation quality in ion beams, especially signal quenching in high-LET beams [6]. In turn, their secondary signals (such as LET-dependent ratios between peaks in a glow curve or OSL emission bands), can be utilized for radiation quality determination and response correction [7]. It seems, however, hardly feasible to establish a unique relation between these secondary signals and a single number specifier, let alone a spectrum, for ions heavier than protons [8].

Powerful detector setups to characterize radiation quality in great detail , i.e. phase space, energy- or microdosimetric spectra, feature mostly active, silicon-based devices [9, 10, 11]. Small, passive solid-state track-detectors such as track etch detectors (TEDs) or fluorescent nuclear track detectors (FNTDs) can usually not match those setups in accuracy, energy resolution or sensitivity but they allow for greater simplicity and different kinds of application, such as in-situ use or high dose-rate exposures.

In these track detectors, assessment of track density and energy loss on a single track level enables to determine $\Phi_L = d\Phi(L)/dL$. Due to the small average range of secondary electrons from ion interactions, dose can also ultimately be approximated by

$$D = \sum_i \frac{L_i}{\rho} \cdot \Phi_{L_i} \quad (1)$$

with $\rho$ being the mass density [12, 13, 14].

Relatively novel alumina-based FNTDs show higher spatial resolution and better sensitivity for low-LET particles than TEDs, which should allow for the recording of higher fluences and the detection of the entire particle spectra found in clinical ion beams (REF Osinga). In addition, alumina single crystals show good biocompatibility. These properties make FNTDs candidates for a number of applications in IBCT [15] with respect to dosimetry [16], radiobiology [17] and treatment verification.

Studies have shown the principle suitability of FNTDs to assess $\Phi_L$ in therapeutic carbon beams [18, 14], but required a considerable amount of manual analysis, lack of statistics, and raised doubt on the detectability of light fragments, esp. protons. Here, we investigated therefore the performance of alumina-based FNTDs for fluence, energy loss, and dose determination at distinct positions of a spread-out carbon ion Bragg peak (SOBP) using a mostly automated reader device and analysis.

## 2. Material and Methods

### 2.1 Fluorescence nuclear track detectors

Alumina single crystals doped with carbon and magnesium ($Al_2O_3$:C,Mg) contain a large density of aggregate $F_2^{2+}$(2 Mg) defects. These color centers change can into $F_2^+$(2 Mg) upon ionizing radiation. The $F_2^+$(2 Mg) centers exhibit intra-center fluorescence with both a broad excitation- (centered at 620 nm) and emission-band (at 750 nm). The short fluorescence life-time (75+/-5 ns) enables confocal laser scanning microscopy. Since the fluorescence intensity is related to the pattern of energy deposition, single ion tracks and their respective energy loss can be recorded [14].

All detectors used in this study were produced by the Crystal Growth Division of LANDAUER Inc. (Stillwater, Oklahoma, USA) and cut along the optical c-axis into small rectangular plates (4.0×8.0×0.5 $mm^3$). One of their large sides is polished to optical quality to enable imaging.

## 2.2 Irradiation

FNTDs were irradiated at the Heidelberg Ion-Beam Therapy Center (HIT) in a spread-out Bragg peak ranging from approx. 10 to 15 cm depth in water. The SOBP was optimized for homogenous biological dose and composed of 18 pristine carbon Bragg peaks with energies between 219 and 280 MeV/u. FNTDs were placed at seven distinct positions in the entrance channel, SOBP and tail, using slabs of RW3 (PTW Freiburg) water equivalent plastic. Taking into account the water-equivalent thickness (WET) of the beam application system and the distance to the iso-center (0.3 mm [19]), and the thickness of the FNTDs that were facing with their non-polished side towards the beam (0.165 cm WET), the effective depths were 0.75, 1.75, 8.75, 10.75, 12.75, 14.75 and 17.75 cm.

The planned entrance fluence of approx. $1.54 \cdot 10^7$ / $cm^2$, corresponding to a 2 GyRBE in the SOBP was scaled down by a factor of 10 (yielding $1.54 \cdot 10^6$ / $cm^2$ entrance fluence) in order to stay well below the fluence limits of the FNTDs even in the presence of fragments.

## 2.3 Monte Carlo transport simulation

Reference data for the irradiations were created using the FLUKA code, version 2011.2c.5 [20, 21]. The same settings were as in [8] were used. In brief, HADROTHE defaults and the DPMJET-III and RQMD libraries were employed, and the Sternheimer parameters and ionization potential (76.8 eV) for water were adjusted. Scoring of depth-curves of fluence, fLET, dLET, and dose for the primary ions and all fragments with charge $Z = 1..5$ was done using the USRBIN card together with a customized FLUSCW routine. In addition, the fluence double-differential in LET and polar angle, $d^2\Phi_Z(L,\theta)/dL \cdot d\theta$, was acquired at each sample position using a combination of USRYIELD and AUXSCORE cards. The beam was assumed to exit the accelerator with a slight energy spread (0.2 % FWHM), and the beam application system and ripple filter were represented by a simplified geometry. Results for the depth-dose curve from the Monte-Carlo simulations did match the irradiation plan better than 3 % at the FNTD positions. The irradiation plan relies on basic data from the treatment planning system which are subject to extensive quality assurance.

**2.4 Detector read-out**

In most previous study by our group, ion irradiated FNTDs have been read out by a commercial laser scanning microscope (ZEISS LSM 710 ConfoCor 3) with the configuration described in [15]. Using a high-NA, oil-immersion objective and avalanche photodiodes (APDs) in photon-counting mode, very good image quality could be achieved. Sample processing – and hence track statistics – is, however, hampered by manual operation. Also, early saturation of the APDs limits the usable dynamic range.

The LANDAUER FXR700RG CLSM - a research version of the commercial system [22] - is in contrast fully automated and dedicated to high-throughput processing. It features an air objective (100x/0.95NA), current-mode APDs and 2D galvoscanning. The 640 nm laser diode was measured to provide 4 mW power at the sample (measured using an Ophir NOVA II powermeter). The laser power is significantly higher compared to the LSM710 (approx. 40 times), and the user has to be aware of potential signal bleaching, which we found however to be in the permille range when re-reading image stacks at the same position. The confocal pinhole is fixed to 1 AU, and an APD in current mode is used to detect the fluorescence light. The FXR700RG processes samples in a 'massive scan', i.e. reading multiple stacks across each detector (Fig. 1). For the samples irradiated in this study, 49 (7x7) frames of 100x100 µm² (504x504 pixel) were read-out per sample, with 41 slices in depth (0 – 100 µm in 2.5 µm steps). The time per image was 10 s (pixel dwell time ca. 40 µs).

**2.5 Image processing**

All image processing was done using the 'FNTD' extension package for the R language [23]. It is targeted towards the analysis of swift ion tracks in FNTDs and provides a scriptable interface to the Java-based 'FNTD' library which in turn uses the NIH-hosted image-processing software IMAGEJ [24] as well as a number of its plugins as a class library. For this study, version 1.0.4 ('classic') of the 'FNTD' package was used. It allowed to run the entire processing from raw image data to final spectra in a single script and is provided as open-source software under GPL3 license at http://fntd.dkfz.de.

The package can import the image output of both the FXR700RG (32-bit float text files) and the LSM710 (16-bit integer, Zeiss LSM format). Images were first background-subtracted using a histogram-based routine from the MOSAIC ToolSuite for IMAGEJ [25]. Then, a modified version of the MOSAIC feature point extraction algorithm [26] was used to identify ion tracks. This tool is specifically tailored for identification ("tracking") of ion beam footprints in fluorescence images ("trackspots") and linking them correctly to reconstruct the ion tracks. Its success rate and throughput for FNTD images is significantly higher that of the original version and also makes it capable of successfully processing high-fluence images [27]. As estimator for LET, the mean intensity $\bar{\eta}$ along each track were calculated as described in [28].

While intensity data from the LSM710 have to be corrected for variable laser-power and saturation effects [14], the output from FXR700RG can be used directly. Corrections for sensitivity fluctuation across the detector area, spherical aberration, field-of-view non-uniformity, and angular dependence of the fluorescence signal were applied as described by [18].

**2.6 LET calibration and assessment**

A relation between the energy loss of a particle $L$ and fluorescence intensity $\bar{\eta}$ was established with the same set of 66 FNTDs as in [14], but read out on the FXR700RG. The set covers irradiations with protons, helium, carbon and oxygen ions and an LET range in alumina from approx. 1 to 150 keV/µm. While sampling approximately twice the volume (four stacks per detector), the read-out of all samples took less than 24 hours in total compared to about three person weeks (24 h / 7 days a week) for the LSM710. A logarithmic function slightly modified to the one used in [14],

$$\bar{\eta} = a \cdot \log\left(\frac{L_{Al_2O_3}}{b} + 1\right) \qquad \text{Eq. 4}$$

with $a = 3.589$ and $b = 5.693$ (Fig. 3) was fit to the data. LET could then be evaluated from fluorescence intensity $\bar{\eta}$ by

$$L_{Al_2O_3} = b \cdot \left(10^{\frac{\bar{\eta}}{a}} - 1\right) \qquad \text{Eq. 5}$$

The function is purely empirical and the underlying mechanism still unknown. Alas, some properties and parameters can be interpreted:

- Intensity response to energy loss is not linear, but considerably sublinear. This is also the case when using the sum of pixel intensities rather than the maximum to derive $\bar{\eta}$.
- The function is designed to approach $L \rightarrow 0$ for $\bar{\eta} \rightarrow 0$, but does not saturate for high LET values.
- $a$ is a scaling factor, accounting for the optical system, the processing of light counts, but especially for the sensitivity of the corresponding FNTD(s).
- $\log b$ represents the slope of the intensity-log LET-relation for large $\eta$ and $L$ values.

$b$ was assumed to be the same for all FNTDs used in the SOBP measurements. $a$, however, can vary due to coloration fluctuation. To determine $a$ for each FNTD, the fluence-weighted LET from transport simulation was matched for the high-LET peak (first five depths) or the low-LET peaks (other depths), respectively.

**2.7 Fluence and dose assessment**

The mean intensity $\bar{\eta}$ spectra (Fig. 3) were then used to divide particles into three groups: primary particles, low-LET and high-LET fragments. The fluence of a single particle $i$ assigned to a group $j$, $\Phi_{ij}$, was determined by dividing the measured particle's track lengths $l_i$ by the readout volume $V$ (Fig. 2):

$$\Phi_{ij} = \frac{l_i}{V} \qquad \text{Eq. 6}$$

The fluence of a group (and eventually the total fluence) was calculated by summing the respective particle fluences. Eq. 3 can be used to obtain the dose of particle group $j$ by

$$D_j = \sum_{i=1}^{N_j} \Phi_{ij} \cdot \left(L_{Al_2O_3}\right)_{ij} \qquad \text{Eq. 7}$$

where $N_j$ is the total number of tracks in group $j$ and $\Phi_{ij}$ the fluence and $\left(L_{Al_2O_3}\right)_{ij}$ the LET of the individual track. The total dose is given by $D = \sum_j D_j$.

## 3. Results and discussion

**3.1 LET spectra**

Fig. 2 shows the experimental results for $\Phi_L$ together with the reference data from the Monte Carlo transport simulation. Distinct peaks appear that can be well separated and attributed to primary carbon ions and secondary fragments. The measured peak for the carbon primaries appears wider than in the reference (most pronounced in the entrance channel) which reflects the linewidth of the FNTD system. In a previous study, values for the relative linewidth of 3.8 to 6.5 % (low-LET limit) were found [28]. Using Eq. 5, this translates into a corresponding relative broadening of 14 % (at 100 keV/μm) to 4 % (1 keV/um). Together with the widening of the initially narrow energy distribution by straggling and fragmentation, the impact of linewidth is less obvious at lower LET and larger depths.

The variation in the sensitivity parameter $a$ obtained from peak matching was found to agree with values in previous studies (approx. 20 %, 1 s.d., [28]). The procedure itself, however, introduces a problematic amount of arbitrariness, which is amplified by the power-law dependence of LET on intensity.

From the distal edge of the SOBP (14.75 cm) on, no carbon ion peak was detected in the FNTD as expected in the tail region. To a small extend, LET values larger than expected for the carbon ions were recorded. Rather than short-range heavy fragments, they most likely represent artifacts caused using the LET-intensity-relation outside its calibration range (i.e. above 150 keV/µm). Since those high-LET events can – even at very low fluence – significantly influence the results, they were excluded from later LET and dose evaluation. Medium-LET (lithium, beryllium, or boron) peaks are visible at SOBP depths, especially with optimized scaling. Protons (represented by the lowest LET peak) are largely underestimated at all positions and only significantly present at the two largest depths.

**3.2 Fluence, fluence-weighted LET, and dose**

The latter results in a considerable underestimation of fragment and total fluence, while the fluence of primary ions agrees well with the reference data (Fig. 4, Tab. 1). Fragment data, however, matches the reference considerably better when protons are excluded ($Z > 2$). The absence of low-LET protons is also reflected in the measured fluence-weighted LET with values for carbon ions showing the same trend as the reference but the LET for fragments being significantly overestimated. Consequently, measured dose values are slightly underestimating the reference. As lighter fragments contribute less to the total dose in the entrance part and the SOBP, the underestimation seen in the fluence is mitigated for dose.

**3.3 Underestimation of light fragment fluence**

The significant underestimation of especially the proton fluence seemingly contradicts findings that show the ability of the alumina-based fluorescent nuclear track detector to reliably record particles with LET values lower than 0.3 keV/µm (in water) [29, 30].

Visual inspection of the FXR raw image data and tracking results did however not reveal a considerable number of trackspots that could be seen by eye and had simply not been identified by the routine and excluded from analysis. Thus, to assess the impact of the inferior optical performance of the FXR reader compared to the LSM 710 on track detection, FNTDs irradiated at mid-SOBP position (i.e. with complex particle spectrum) were read-out with both two systems. The resulting images show a similar number of low-LET tracks with large polar angle when read-out with standard settings (Fig. 4). Maximizing the performance of the LSM beyond acquisition times reasonable for standard analyses does reveal additional tracks - however, those tracks are very faint and hardly to detect by current image processing. This is also reflected by the fact that the quantitative analysis of the track density yielded no considerable difference between the FXR and LSM (regular) read-out, and an increase of approx. 30 % in tracks for the maximum performance settings. They can however not explain the factor of three needed to match the reference data.

As shown in Fig. 5, fragments with polar angles larger than approx. 7.5 degrees are not found in the experimentally determined spectrum. In contrast, the vast majority of protons but also a considerable fraction of helium ions are predicted to be produced at larger polar angles in inelastic nuclear scattering events. At higher polar angles, trackspots in the FNTD images become elongated and the lateral distance between trackspots from the same track with depth increases. This deteriorates the ability of the image processing to identify tracks. In addition, the fluorescence signal drops considerably with larger polar angles [18, 14]. This can cause low-intensity trackspots to fall below the detection threshold as shown in Fig. 6. Both effects are thus connected most likely aggravated in the presence of bright trackspots from the primary carbon ions. The estimates given in Fig. 7 indicate that the discussed effects are responsible for the underestimation of fluence of light fragments.

## 4. Conclusion

A semi-automated FNTD workflow can be used for fluence and LET assessment in a spread-out Bragg peak from carbon ions at approx. 1/10 of the clinically used dose. The user interaction is limited to the identification of suitable parameters for trackspot identification and linking, and sensitivity calibration of the individual FNTDs. Despite the inferior optical specifications of the FXR700RG reader compared a high-NA, oil-immersion system like the Zeiss LSM710, no disadvantage for the presented application was seen. Rather, better dynamic range in light detection is crucial for carbon beams and high-throughput capability not only enables to achieve good track statistics and a reasonable number of samples per experiment but also enables more accurate corrections such as spherical aberration or field-of-view non-uniformity due to the large amount of data.

While the fluence of primary particles can be reliably detected and their LET spectrum determined in relative terms, the assignment of absolute LET values is severely hampered by the fluctuation in sensitivity of the individual FNTDs. It remains to be seen, if more homogenous crystals or suitable coloration correction methods will improve the situation. In addition, the use of LET-related quantities less dependent on crystal sensitivity might help [31].

The significant underestimation of low-LET particle fluence is mainly caused by the wide angular distribution for secondary protons in heavier ion beams. Since the detection of particles with large polar angles seems generally challenging, it is at least questionable if better sensitivity and spatial resolution will have a considerable impact. From an energy deposition point of view, however, light fragments do not match primary and heavier ions in importance. Also, in protons (and most likely also helium) beams, the detection success rate is better due to a much narrower angular spectrum and the absence of bright trackspots [32]. Thus, FNTDs still represent a valuable tool to determine the particle spectrum in clinical ion beams.

**Acknowledgements**

The authors would like to thank Prof. Mark Akselrod for his longstanding support of our project. We greatly benefitted from the most competent and uncomplicated help for the FXR reader by Jonathan Harrison and Vasiliy Fomenko (Landauer Inc.). Our studies would not have been possible with the help of many colleagues at the Heidelberg Ion Therapy Center (HIT), most prominently we gratefully acknowledge Dr. Andrea Mairani's help with the FLUKA code, and Dr. Stephan Brons being an invaluable support before, during, and after beam time shifts.

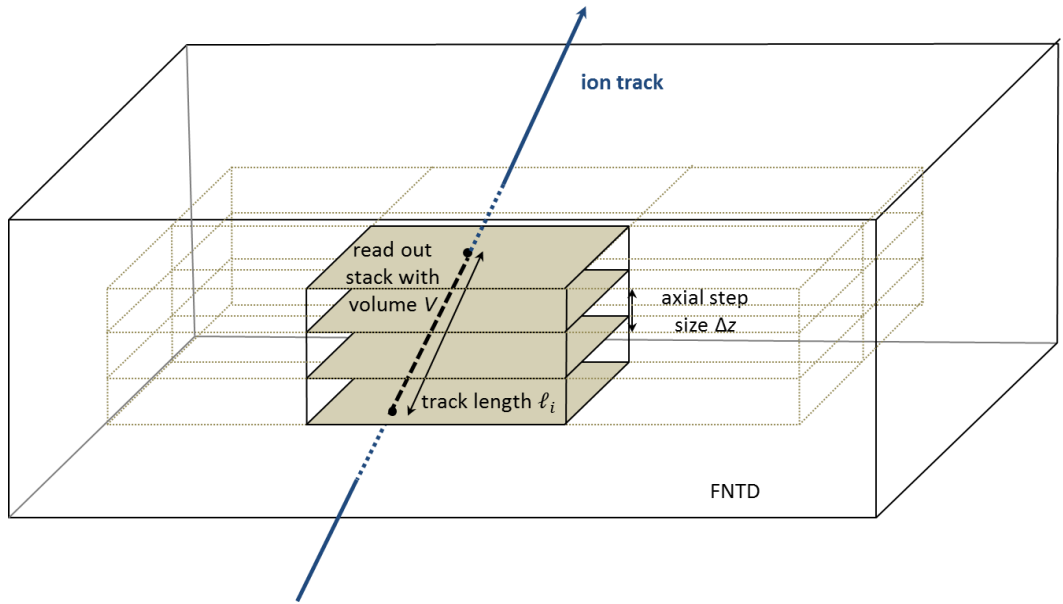

*Figure 1: Cartoon of an FNTD readout. A single track crosses with a length $l_i$ the read out volume $V$ of a stack with four slices in depth. In total, six stacks are read in the FNTD arranged as three by two frames in this example.*

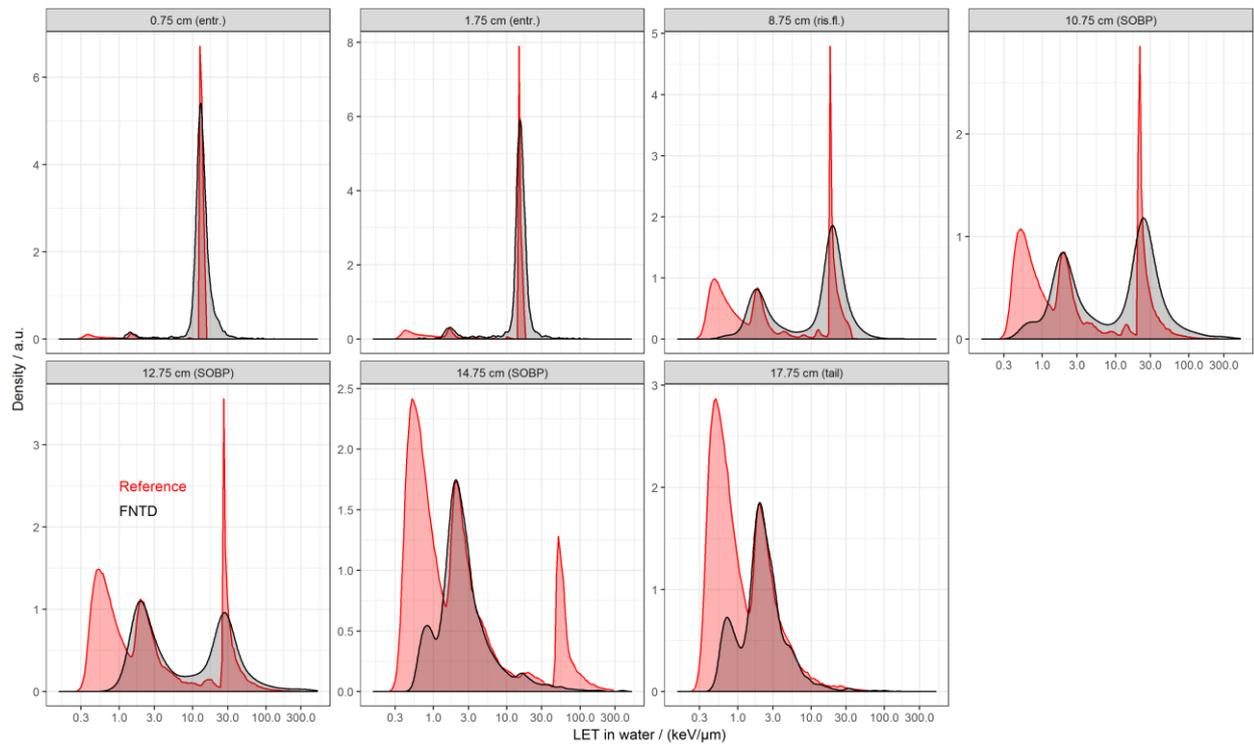

*Figure 2: LET distributions $\Phi_L$ for the seven measurement depths. Reference data from transport simulation are given in red, measured data in black. The data were scaled for the second (He) peak to have the same height.*

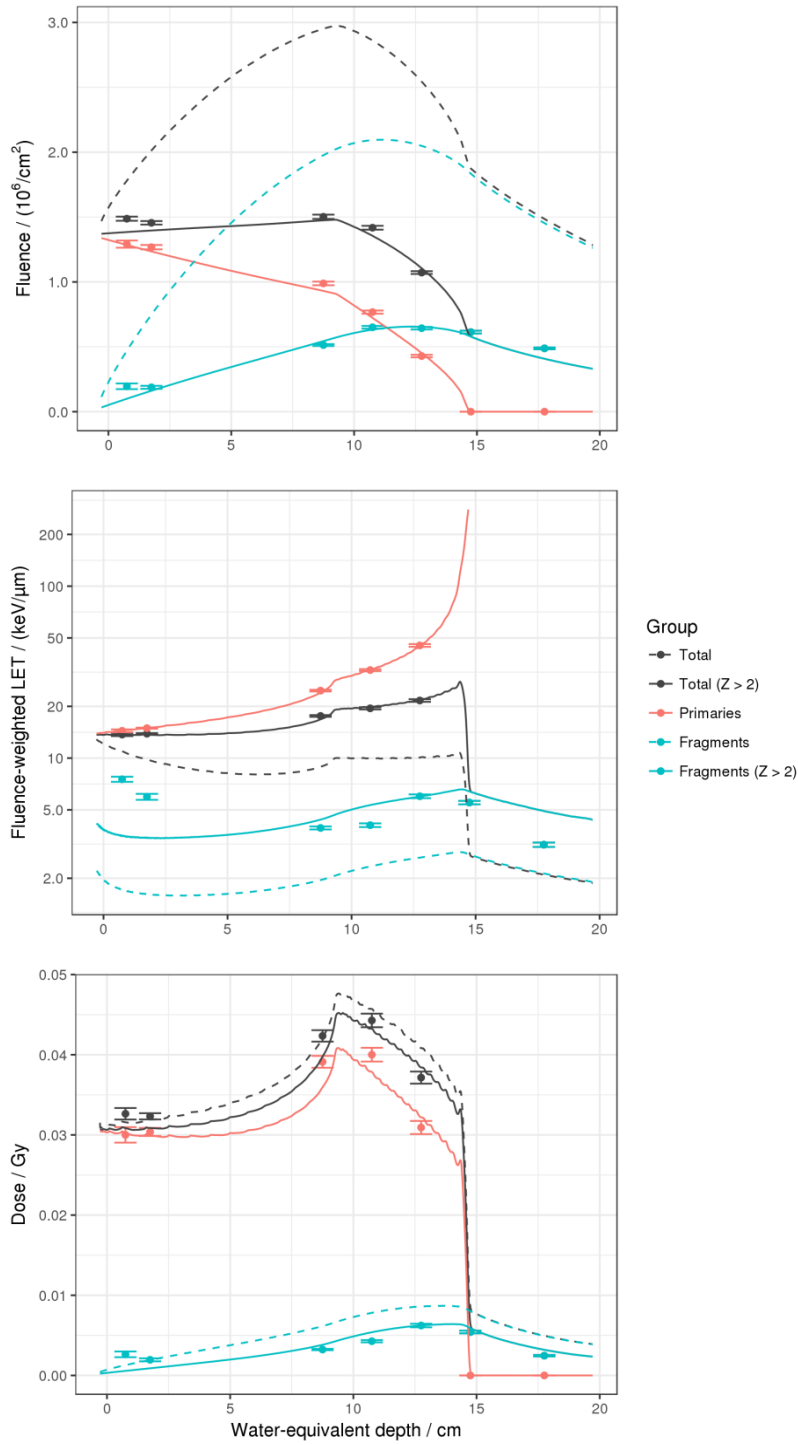

*Figure 3: Reference and measured data for fluence, fluence-weighted LET, and dose. A fluence of $1 \cdot 10^6$ cm$^{-2}$ corresponds to 100 tracks measured per frame and 4900 tracks per sample, respectively.*

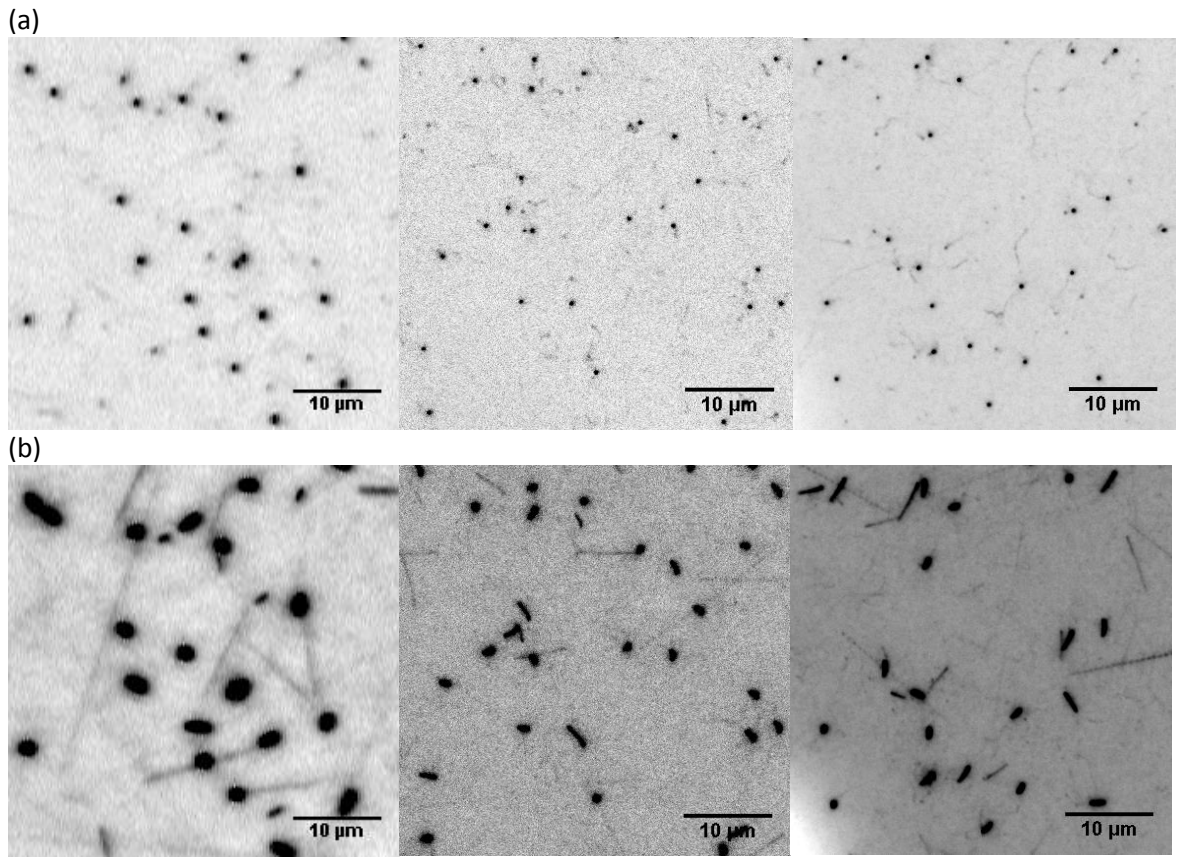

*Figure 4: (a) Fluorescence images of an FNTD located at the mid-SOBP position, slice at 5 μm depth read-out using the Landauer FXR700RG (left column), and the Zeiss LSM710 ConfoCor 3 with regular (middle) and optimized settings (right). For the latter, the pixel dwell time was doubled and the four read-out were averaged, resulting in an acquisition time of 10 min per image (10 s for the FXR, 1 min for the LSM regular settings). The lower NA of the objective lens of the FXR700RG results in a lower spatial resolution (FWHM) of 0.36 μm (1.7 μm) lateral (axial) in contrast to 0.20 μm (0.8 μm) for the LSM710 in the configuration used. Row (b) shows an average intensity projection of the corresponding images in (a), illustrating the polar angles of the ion tracks and enhancing low-LET, considerably oblique tracks. A projected track length of 10 μm corresponds to a polar angle of approx. 5.5 (10) degree for the FXR700RG (LSM710).*

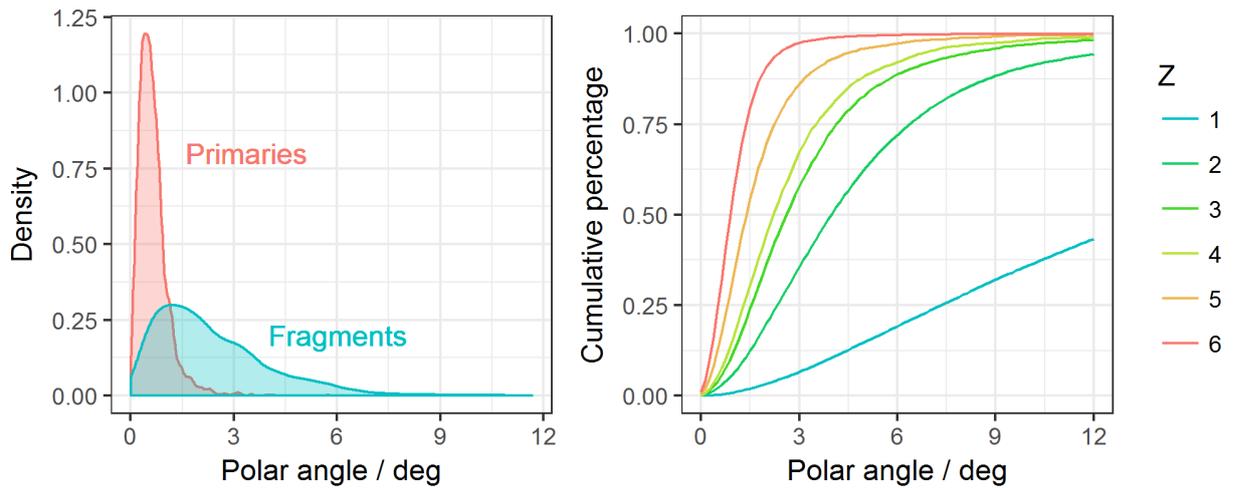

*Figure 5: Left - Measured polar angle distribution at mid-SOBP position (depth: 12.75 cm) for primary carbon ions (red) and fragments (blue). Right – Cumulative percentage of ions as a function of polar angle and charge Z as given by FLUKA for the mid-SOBP position. While carbon ions show a narrow distribution, only a fraction of the light fragments, esp. protons, are found within 10 degrees. Due to the minor lateral scattering of the primary ions and the wide angular distribution for protons in the fragmentation process, the picture for other depth does not change dramatically.*

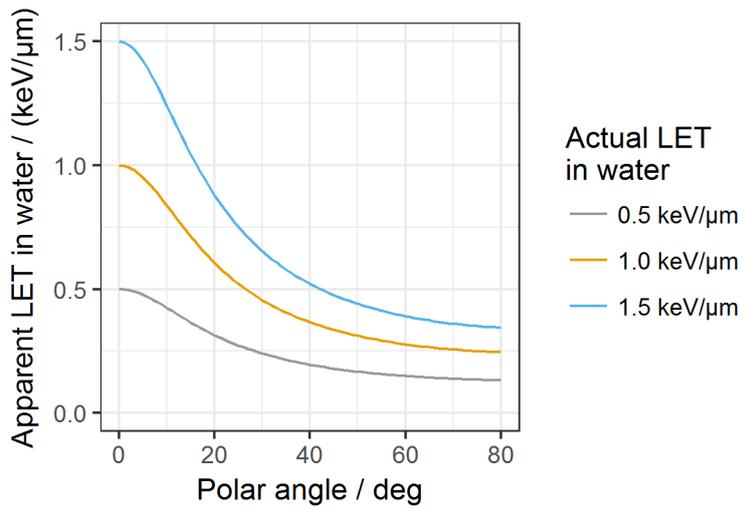

*Figure 6: Decrease of apparent LET due to lower intensity of trackspots with increasing polar angle. At approx. 30 degrees, the intensity has dropped to 50 %. Despite the logarithmic dependency on intensity, the apparent LET response similarly.*

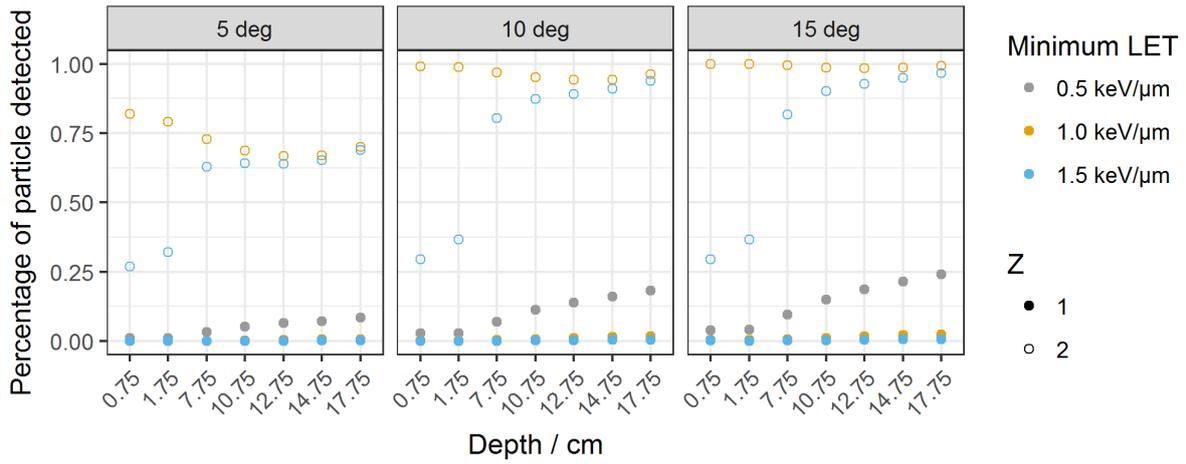

*Figure 7: Proton and helium fragments detection ratios derived from the FLUKA simulation for three minimum LET- and three maximum polar angle thresholds, respectively. For 0.5 and 1.0 keV/μm, no difference is found for helium (since LET is always > 1.0 keV/μm) and symbols fall on top of each other. Except for shallow depths (where low numbers of fragments occur), the minimum LET threshold is most important for protons and the maximum polar angle (between 5 and 10 degrees) influences considerably the helium detection.*

| Position / cm | Component | Fluence / $10^6$ cm$^{-2}$ | fLET Water/ (keV/µm) | Dose / mGy |
|---|---|---|---|---|
| 0.75 | Low-LET fragments | 0.20 (0.02) | 7.5 (0.2) | 2.6 (0.4) |
| | Primaries | 1.29 (0.02) | 14.3 (0.3) | 30.0 (1.0) |
| | Total | 1.49 (0.03) | 48.8 (2.5) | 32.6 (0.7) |
| | High-LET fragments | 0.010 (0.002) | 71.3 (3.4) | 1.2 (0.2) |
| 1.75 | Low-LET fragments | 0.19 (0.01) | 6.0 (0.1) | 2.0 (0.2) |
| | Primaries | 1.27 (0.02) | 14.9 (0.1) | 30.4 (0.5) |
| | Total | 1.46 (0.01) | 13.9 (0.1) | 32.3 (0.4) |
| | High-LET fragments | 0.014 (0.002) | 64.9 (2.5) | 1.6 (0.2) |
| 7.75 | Low-LET fragments | 0.51 (0.01) | 3.9 (0.1) | 3.2 (0.08) |
| | Primaries | 0.99 (0.01) | 24.7 (0.3) | 39.1 (0.7) |
| | Total | 1.50 (0.02) | 17.6 (0.2) | 42.3 (0.7) |
| | High-LET fragments | 0.007 (0.001) | 252 (36) | 2.7 (0.6) |
| 10.75 | Low LET fragments | 0.65 (0.01) | 4.1 (0.1) | 4.3 (0.1) |
| | Primaries | 0.77 (0.01) | 32.5 (0.4) | 40.0 (0.7) |
| | Total | 1.42 (0.01) | 19.5 (0.3) | 44.3 (0.8) |
| | High-LET fragments | 0.017 (0.002) | 543 (32) | 15.1 (1.9) |
| 12.75 | Low LET fragments | 0.64 (0.01) | 6.0 (0.2) | 6.2 (0.2) |
| | Primaries | 0.43 (0.01) | 45.2 (0.9) | 30.9 (0.8) |
| | Total | 1.07 (0.01) | 21.6 (0.4) | 37.1 (0.8) |
| | High-LET fragments | 0.022 (0.002) | 770 (68) | 25.7 (2.8) |
| 14.75 | Low LET fragments | 0.61 (0.01) | 5.5 (0.1) | 5.4 (0.2) |
| | High-LET fragments | 0.008 (0.001) | 202 (21) | 2.5 (0.4) |
| 17.75 | Low LET fragments | 0.49 (0.01) | 3.1 (0.1) | 2.5 (0.1) |
| | High-LET fragments | 0.0008 (0.0004) | 280 (45) | 0.4 (0.3) |

*Tab. 1: Results for fluence, fluence-weighted LET, and dose, given for the tracks assigned to the group of primary particles, low- and high-LET fragments, respectively. In brackets, the standard uncertainty is given as evaluated from the variance between the frames. For the two largest depths, only low-LET fragments were assumed to be recorded. The actual existence of "high-LET" fragments cannot be definitely concluded from the spectra and has to be discussed with respect to the deficiencies of the used logarithmic intensity-LET-relation.*